%% file: main.tex
\documentclass{article}

\usepackage[accepted]{icml2026}

\usepackage[utf8]{inputenc}
\usepackage[T1]{fontenc}
\usepackage{microtype}          

\usepackage{amsmath, amssymb, amsthm}
\usepackage{bm}                 

\usepackage{tikz}
\usetikzlibrary{arrows.meta, decorations.pathreplacing, calc, shadings}
\usepackage{standalone}
\usepackage{graphicx}
\usepackage{booktabs}
\usepackage{array}
\usepackage{multirow}
\usepackage[caption=false,font=footnotesize]{subfig}
\graphicspath{{figures/}}

\usepackage[colorlinks=true, linkcolor=blue, citecolor=blue, urlcolor=blue]{hyperref}
\usepackage{acro} 
\usepackage{xcolor}
\usepackage{cleveref}
\usepackage[inline]{enumitem}

\PassOptionsToPackage{hyphens}{url}\usepackage{hyperref}
\usepackage{xurl}  

\icmltitlerunning{Single-Cell Cross-Modal Transfer by Adversarial Fine-Tuning of Foundation Models}

\date{}   

\input{acronyms}

\begin{document}

\twocolumn[
  \icmltitle{Single-Cell Cross-Modal Transfer by Adversarial Fine-Tuning of Foundation Models}

  \begin{icmlauthorlist}
    \icmlauthor{Joseph Boyd}{gsk}
    \icmlauthor{Matthew Lyon}{gsk}
    \icmlauthor{Martino Mansoldo}{gsk}
    \icmlauthor{Christian Hurry}{gsk}
    \icmlauthor{Finnian Firth}{gsk}
  \end{icmlauthorlist}

  \icmlaffiliation{gsk}{GSK.ai, GlaxoSmithKline, London, United Kingdom}

  \icmlcorrespondingauthor{Joseph Boyd}{joseph.x.boyd@gsk.com}

  \icmlkeywords{spatial transcriptomics, foundation models, adversarial learning, cross-modal transfer}

  \vskip 0.3in
]

\printAffiliationsAndNotice{}

\begin{abstract}
  \input{sections/abstract}
\end{abstract}

\input{sections/introduction}
\input{sections/related_work}
\input{sections/method}
\input{sections/experiments}
\input{sections/results}

\input{sections/conclusion}

\section*{Acknowledgements}
We extend our thanks to Lefteris Zormpas, Stefan Groha, and Richard Li for providing helpful feedback and discussions.

\section*{Impact Statement}
The objective of this work is to better understand the connections between different single-cell modalities, in an effort to better understand fundamental biology and characteristics of disease revealed by gene expression. This work could therefore have a general scientific impact as well as an impact on the development of future therapies.

\bibliography{references}

\appendix
\input{sections/appendix}

\end{document}

%% file: acronyms.tex
\DeclareAcronym{st}{short=ST,long=spatial transcriptomics}
\DeclareAcronym{scrnaseq}{short=scRNA-seq,long=single-cell RNA sequencing}
\DeclareAcronym{emd}{short=EMD,long=earth mover's distance}
\DeclareAcronym{wgan-gp}{short=WGAN-GP,long=Wasserstein GAN with gradient penalty}
\DeclareAcronym{atacseq}{short=ATAC-seq,long=assay for transposase-accessible chromatin using sequencing}
\DeclareAcronym{hvg}{short=HVG,long=highly variable gene}
\DeclareAcronym{kr}{short=KR,long=Kantorovich-Rubinstein}
\DeclareAcronym{gsea}{short=GSEA,long=gene set enrichment analysis}
\DeclareAcronym{pcc}{short=PCC,long=Pearson correlation coefficient}

%% file: sections/abstract.tex
\Ac{st} is a powerful tool for exploring biological properties dependent on structure, proximity, and interaction in tissue. The methods underpinning \ac{st} are developing rapidly but are limited in their ability to profile many thousands of genes at a subcellular scale. Although dissociated from tissue, it is known that the whole-transcriptome readouts of cells in \ac{scrnaseq} retain information about their former \emph{in situ} neighbourhoods, motivating computational methods to recover it. While paired \ac{st} and \ac{scrnaseq} datasets are scarce, each modality in its own right is abundantly available. We therefore propose to perform cross-modal translation between unpaired \ac{st} and \ac{scrnaseq} data. In this work we show that a single-cell foundation model can perform this translation via adversarial fine-tuning. We demonstrate that our method performs favourably against methods built for multi-omics translation.

%% file: sections/introduction.tex
\section{Introduction}
\label{sec:introduction}

Methods for spatial transcriptomics (\ac{st}) provide spatially resolved readouts of gene expression in biological samples. In particular, fluorescence \emph{in situ} hybridisation (FISH) methods such as MERFISH~\cite{chen2015spatially} and Xenium~\cite{janesick2023high} provide subcellular resolution to spatial biology, but are often limited in their coverage of the transcriptome. By contrast, single-cell RNA sequencing (\ac{scrnaseq}) modalities provide whole-transcriptome readouts, but for cells that have been dissociated from their original tissue context. Due to these differing strengths, \ac{st} and \ac{scrnaseq} are currently seen as complementary modalities. However, capturing both modalities can be prohibitive in terms of cost and sample availability. Therefore, there is a need for computational methods that can translate between these modalities, allowing one to infer spatial context from \ac{scrnaseq} data alone. Furthermore, such a translation may provide insight into the relationship between these modalities, and the different snapshots of biological state they encode.

\begin{figure*}[htbp]
  \centering
  \subfloat[Cell types\label{fig:foxj1-a}]{%
    \includegraphics[width=0.24\textwidth]{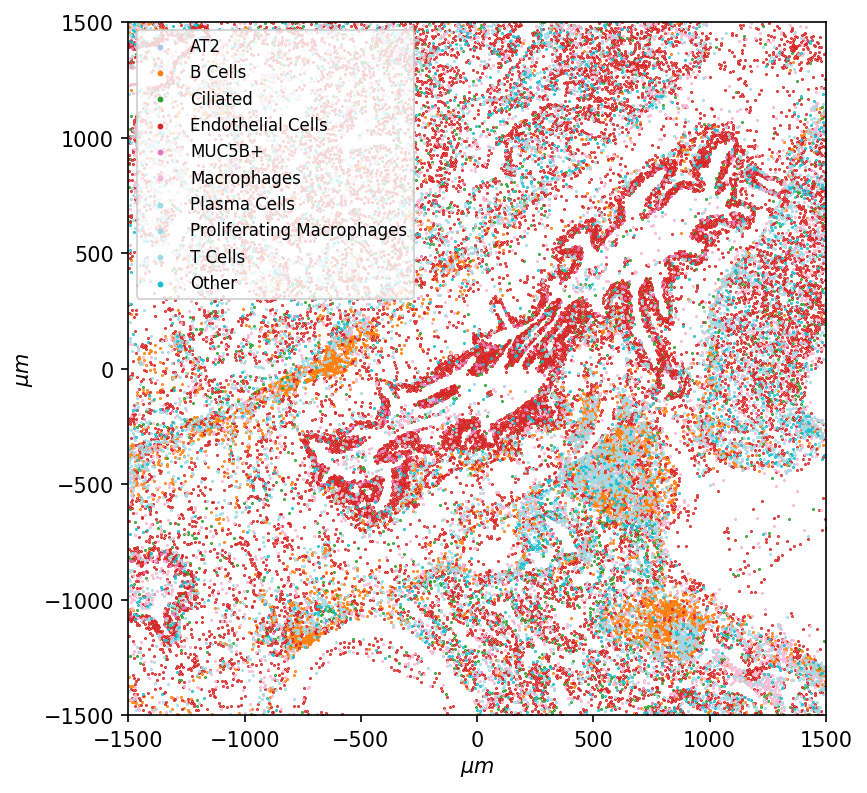}%
  }\hfill
  \subfloat[Gene expression (binarised)\label{fig:foxj1-b}]{%
    \includegraphics[width=0.24\textwidth]{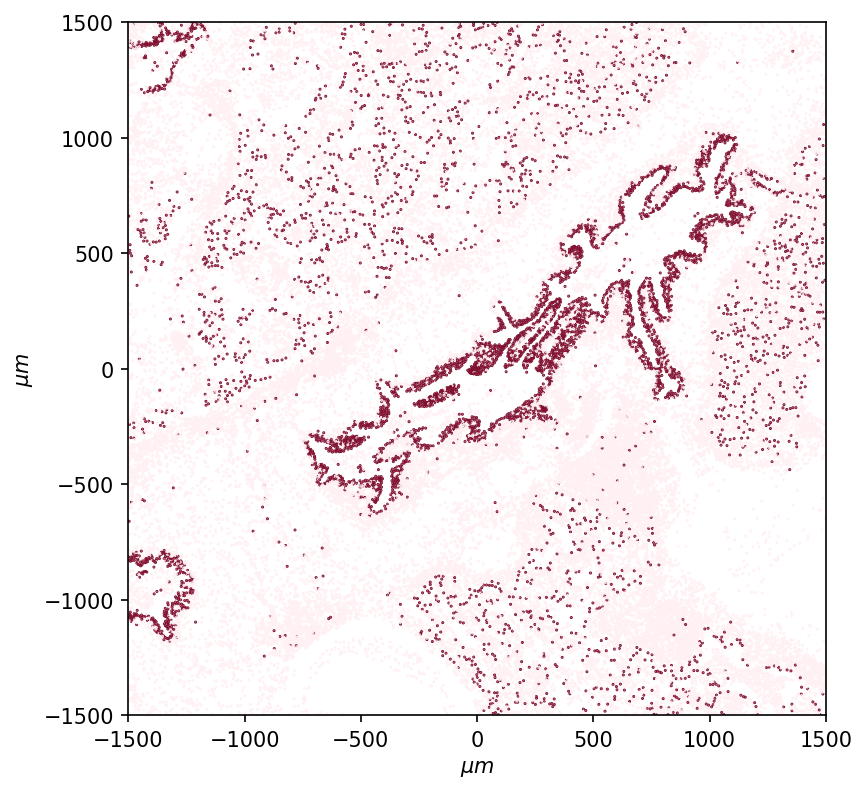}%
  }\hfill
  \subfloat[Spatially bulked\label{fig:foxj1-c}]{%
    \includegraphics[width=0.24\textwidth]{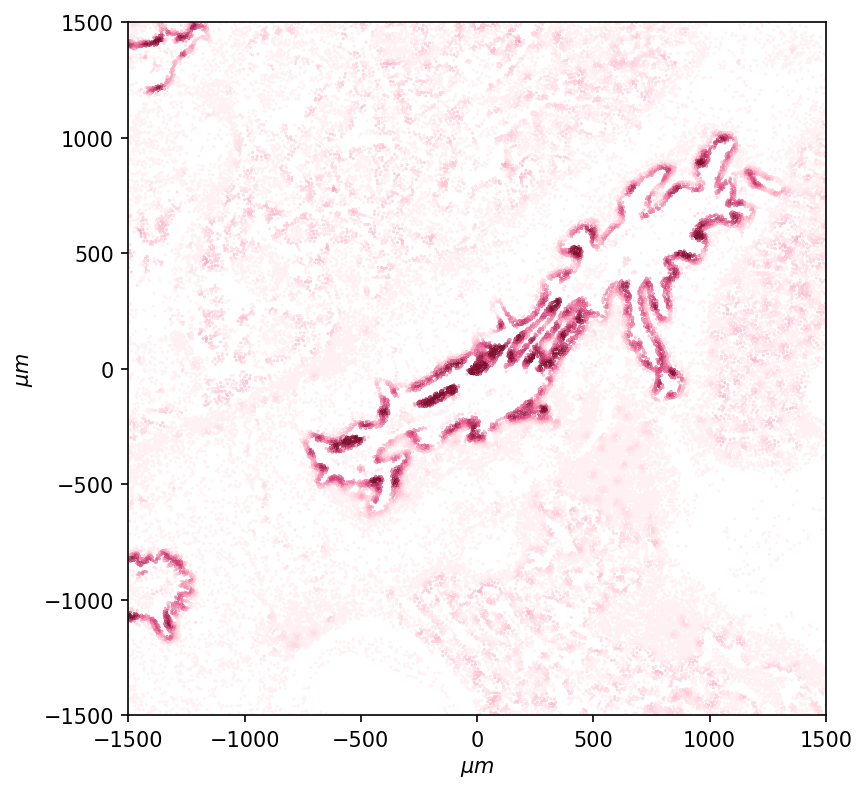}%
  }\hfill
  \subfloat[Model prediction\label{fig:foxj1-d}]{%
    \includegraphics[width=0.24\textwidth]{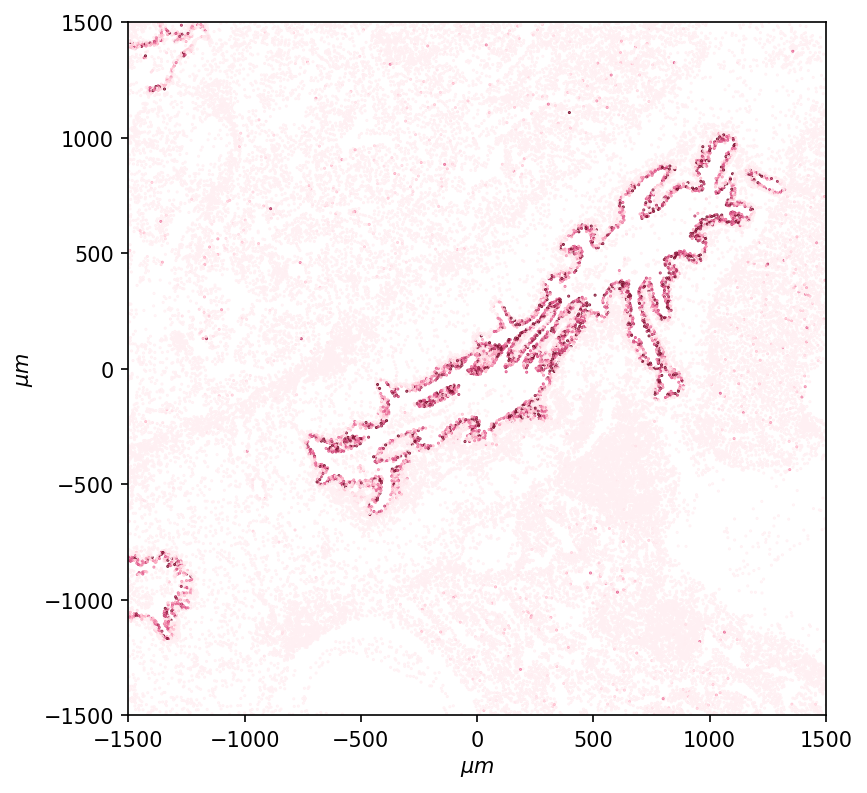}%
  }
  \caption{SCXM identifies single cells with high spatially bulked expression. Training is performed on dissociated single-cell data from different cell types (a), and performance is validated on high-plex spatial data, as depicted with: input gene expression for the \emph{FOXJ1} gene (b), target spatially bulked expression (c), model predictions (d). It is seen that regions of dense gene expression and isolated expression in (b) are respectively promoted and attenuated in (c).}
  \label{fig:foxj1}
\end{figure*}

The aim of this work is to learn to translate, on the level of a single cell, \ac{scrnaseq} gene expression to spatially bulked expression deriving from \ac{st}. The challenge is to learn to perform this translation from fully unpaired data, and the key insight is to exploit the fact that both modalities manifest as count data. We present single-cell cross-modal (SCXM), an adversarial fine-tuning approach for single-cell foundation models aimed at modality translation from \ac{scrnaseq} to \ac{st} context features (Figure~\ref{fig:foxj1}). In so doing, we demonstrate that the gene expression of dissociated cells retains clues as to its former spatial context~\cite{armingol2021deciphering}. Although the strength of this connection varies by gene, we are able to rationalise biologically the relative success of certain pathways.



%% file: sections/related_work.tex
\section{Related Work}
\label{sec:related_work}

The problem of unpaired translation has previously been explored for omics data. scConfluence~\citep{samaran2024scconfluence} used optimal transport to determine likely pairings across modalities, then trained encoders to align the latent embeddings based on these pairings. In contrast, scACT~\citep{xu2024scact} trained encoders that map between modalities, namely \ac{scrnaseq} and ATAC-seq, enforced via a cycle consistency loss~\citep{zhu2017unpaired}. CellContrast~\citep{li2024cellcontrast} designed a method for the recovery of a pseudo-space for single-cell data after contrastive learning on \ac{st} data. Spatial context is often modelled as an aggregation of expression values over the graph connecting neighbouring cells, for example by~\citet{hu2021spagcn, birk2025quantitative, wang2025scgpt}. In contrast to the aforementioned methods, SCXM is trained to infer spatial context from single-cell data directly.

Generative pretraining of large transformer networks has acquired significant currency in omics analysis, following on from its successes in natural language processing and computer vision. Such models have come to be regarded as \emph{foundational}, given their aptitude for few-shot transfer to canonical single-cell applications, including cell typing, batch integration, and perturbation analysis~\citep{yang2022scbert, cui2024scgpt, hao2024large}. Geneformer~\citep{theodoris2023transfer} is a pioneering transformer for inference over transcriptomics data, combining a BERT~\citep{devlin2019bert} transformer architecture with a non-parametric rank value encoder to order genes. The ordered \texttt{<gene\_id>} embeddings combine with positional embeddings to impose a rank on genes, ordered by normalised expression. Nicheformer~\citep{tejada2025nicheformer} extends Geneformer by joint training on spatial and non-spatial data.

%% file: sections/method.tex
\section{Method}
\label{sec:method}

We denote \ac{scrnaseq} gene expression data $X \in \mathbb{R}^{|\mathcal{C}_{\text{SC}}| \times |\mathcal{G}_{\text{SC}}|}$ and \ac{st} expression data $Y \in \mathbb{R}^{|\mathcal{C}_{\text{ST}}| \times |\mathcal{G}_{\text{ST}}|}$, where $\mathcal{C}_{\text{SC}}$ is the set of cells sequenced under \ac{scrnaseq} with the gene panel $\mathcal{G}_{\text{SC}}$, and $\mathcal{C}_{\text{ST}}$ is the set of cells sequenced by \ac{st} with the gene panel $\mathcal{G}_{\text{ST}}$. In practice, $\mathcal{C}_{\text{SC}} \cap \mathcal{C}_{\text{ST}} = \emptyset$. That is, the cells of each modality originate in different tissue samples, or indeed patients. As such, the problem setting is one of unpaired translation. We only require that $\mathcal{G}_{\text{ST}} \cap \mathcal{G}_{\text{SC}} \neq \emptyset$, which in practice is always satisfied.

Our aim is to train a generative encoder model $G_{\text{E}} : X \to Z$, where $Z$ is some representation of spatial context. Our approach will rely on two key assumptions: dissociated cells broadly preserve their formerly \emph{in situ} gene expression when sequenced with a \ac{scrnaseq} modality; and \ac{st} and \ac{scrnaseq} count data, up to a straightforward normalisation, are interchangeable when measured across the same subset of genes. The first assumption will facilitate the learning of a cross-modal mapping, $G_{\text{E}} : X \to Z$, while the second assumption will be used to condition the encoder towards a consistent posterior, $G_{\text{E}}(X) \sim \mathrm{P}_{Z|X}$.


\subsection{Encoding Spatial Context}
\label{subsec:problem}

Cells in \ac{st} are typically modelled as nodes in a spatial graph~\cite{palla2022squidpy}, in which connectivity is determined on the basis of spatial proximity, either as a fixed number of nearest neighbours, or by a prescribed spatial radius. Given such a spatial graph, various measures exist for quantifying the concentration of gene expression in space, including spatial autocorrelation statistics~\cite{moran1950notes, geary1954contiguity}. A natural approach to define a spatial context $Z$ is via a graph convolution, which for a given cell $i$ and gene $g$ is computed as $z_{ig} = \sum_{j \in N(i)} k_{ij}y_{jg}$, where $N(i)$ is the neighbourhood of cell $i$, including a self-loop. The kernel weights $k_{ij}$ may be apportioned in various ways, including uniform and Gaussian weights. In all experiments we use a Gaussian kernel with a radius of $25\mu m$.

\subsection{Learning Framework}
\label{subsec:approach}

Our adopted learning framework is the \ac{wgan-gp}~\cite{gulrajani2017improved}. Wasserstein GANs~\cite{arjovsky2017wasserstein} use the \ac{emd} as a metric for measuring the divergence between generated and ground truth distributions on data. The intractable \ac{emd} is reformulated using the \ac{kr} duality, yielding the training objective $\min_{G_{\text{E}}}\max_{D}\mathcal{L}_{\text{WGAN}}$, where


\begin{equation}
  \mathcal{L}_{\text{WGAN}} = \mathbb{E}_{z \sim p_Z}\big[D(z)\big] - \mathbb{E}_{x \sim p_X}\big[D\big(G_{\text{E}}(x)\big)\big].
\end{equation}

A discriminator $D$ is trained to estimate (in dual form) the \ac{emd}, which the generator $G_{\text{E}}$ is trained to minimise. To satisfy the \ac{kr} duality, the discriminator must be $K$-Lipschitz continuous, a condition that is enforced by a gradient penalty

\begin{equation}
  \mathcal{L}_{\text{GP}} = \mathbb{E}_{\check{z} \sim P_{\check{Z}}}\big[\big(\Vert\nabla_{\check{z}}D(\check{z})\Vert_2 -1\big)^2\big],
\end{equation}

where $\check{z}$ are linearly interpolated samples of generated and ground truth data. We train a decoder $G_{\text{D}}$ to reconstruct \ac{st} gene expression $Y$ from precomputed $Z$ deriving from the \ac{st} modality, subject to

\begin{equation}
  \mathcal{L}_{\text{ST}} = \mathbb{E}_{y \sim p_Y}\big[\Vert G_{\text{D}}\big(z(y)\big) - y\Vert_2^2\big].
  \label{eq:st_consistency}
\end{equation}

\begin{figure}[t!]
  \centering
  \includegraphics[width=1.0\linewidth]{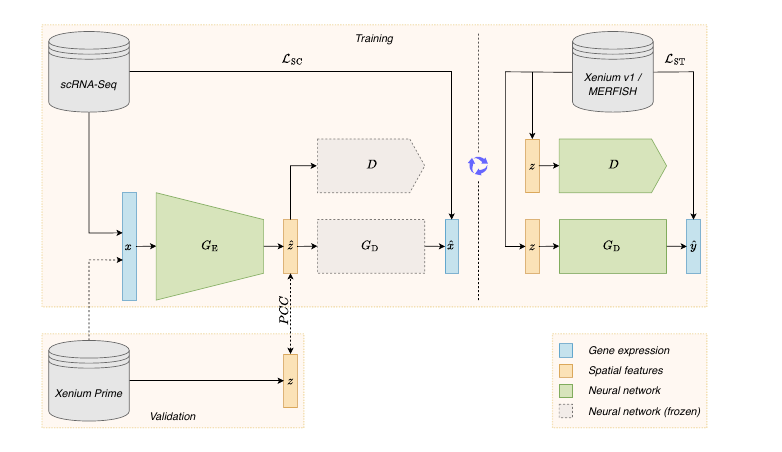}
  \caption{SCXM model architecture, with $G_{\text{E}}$ a fine-tuned foundation model. The adversarial discriminator $D$ enforces that inferred spatial features $\hat{z}$ remain in-distribution by approximating the earth mover's distance, while a decoder $G_{\text{D}}$ enforces a consistent mapping. The decoder and discriminator are trained in alternation with the encoder. Training is performed on fully unpaired data sources, and high-plex spatial data serves as a paired validation set.}
  \label{fig:model-architecture}
\end{figure}

Hence, the encoder $G_{\text{E}}$ can be made consistent by decoding its predicted spatial features with the frozen decoder $\texttt{freeze}(G_{\text{D}})$ by

\begin{equation}
  \mathcal{L}_{\text{SC}} = \mathbb{E}_{x \sim p_X}\big[\Vert \texttt{freeze}(G_{\text{D}})\big(G_{\text{E}}(x)\big) - x_{:\mathcal{G}_{\text{ST}}}\Vert_2^2\big],
  \label{eq:sc_consistency}
\end{equation}

where $x_{:\mathcal{G}_{\text{ST}}}$ denotes \ac{scrnaseq} expression sliced to the spatial panel of genes. In sum, our model resembles an adversarial autoencoder~\cite{makhzani2015adversarial}, trained in a semi-supervised way. The full objective is finally

\begin{equation}
  \min_{G_{\text{E}}, G_{\text{D}}}\max_{D}\mathcal{L}_{\text{WGAN}} +
  \lambda_{\text{GP}}\cdot\mathcal{L}_{\text{GP}} +
  \lambda_{\text{ST}}\cdot\mathcal{L}_{\text{ST}} +
  \lambda_{\text{SC}}\cdot\mathcal{L}_{\text{SC}},
\end{equation}

where $\lambda_{\text{GP}}$, $\lambda_{\text{ST}}$, and $\lambda_{\text{SC}}$ are tunable weights for the respective penalties. The model architecture is shown in Figure~\ref{fig:model-architecture}.

%% file: sections/experiments.tex
\section{Experimental Setup}
\label{sec:experiments}

\subsection{Datasets}
\label{subsec:datasets}

Datasets were curated by combining public \ac{scrnaseq} and \ac{st} datasets according to tissue type. Our data sources were \texttt{CellxGene}~\cite{czi2025cz}, the Gene Expression Omnibus (GEO)~\cite{clough2016gene}, and 10x Genomics preview data (see Extended Table~\ref{tab:datasets}). Given the unpaired learning objective, a key challenge was validation. We approximated a paired validation set (see Figure~\ref{fig:model-architecture}) by using Xenium Prime data~\cite{technical-note-prime}, which is a \ac{st} protocol with a high-plex gene panel $\mathcal{G}_{\text{PR}}$ ($\sim5000$ genes). To ensure cross-modal compatibility, we slice the \ac{scrnaseq} and Xenium Prime data to $\mathcal{G}_{\text{SC}} \cap \mathcal{G}_{\text{PR}}$, and the \ac{st} data to $\mathcal{G}_{\text{ST}} \cap \mathcal{G}_{\text{PR}}$. We describe below the datasets analysed, and their dimensions after filtering and slicing.


\paragraph{Lung Dataset.} The Human Lung Cell Atlas (HLCA)~\cite{sikkema2023integrated} \ac{scrnaseq} ($1,824,518 \ \text{cells} \times 4908$ genes), Xenium V1 pulmonary fibrosis data~\cite{vannan2025spatial} ($1,093,782 \ \text{cells} \times 204$ genes), and Xenium Prime for Human Lung Cancer~\cite{technical-note-prime-lung} ($159,771 \ \text{cells} \times 4908$ genes).

\paragraph{Brain Dataset.} Chromium of mouse brain nuclei~\cite{data-highlights-mouse-brain} ($44,725 \ \text{cells} \times 4980$ genes), MERFISH mouse frontal cortex matter~\cite{allen2023molecular} ($367,582 \ \text{cells} \times 223$ genes), and mouse hemispheric data with Xenium Prime~\cite{technical-note-prime} ($56,799 \ \text{cells} \times 4980$ genes). Mouse genes were mapped to their human homologues for compatibility with the vocabulary of the model backbone using \texttt{BioMart}~\cite{smedley2009biomart}.

\paragraph{Breast Dataset.} A \ac{scrnaseq} atlas of human breast cancer~\cite{chen2026highly} ($404,579 \ \text{cells} \times 5054$ genes), a Xenium V1 breast cancer cohort~\cite{janesick2025biomarker} ($1,275,889 \ \text{cells} \times 148$ genes), and Xenium Prime~\cite{technical-note-prime} of a breast cancer ($107,004 \ \text{cells} \times 5054$ genes).

\subsection{Preprocessing}
To foster consistency between the collected modalities, we performed cell filtering in a harmonised way across the spatial genes $\mathcal{G}_{\text{ST}}$, that is, the genes shared by all modalities. Cell type annotation was performed with CellTypist~\cite{dominguez2022cross} using tissue-appropriate models. Due to the significant domain shift remaining between the modalities, we adopted a simple binarisation binning technique for the reconstruction targets of the decoder $G_{\text{D}}$, that is, an indicator of non-zero gene expression. Graph convolutions were computed on the binarised expression with a customisation of \texttt{scanpy}'s~\cite{wolf2018scanpy} metrics module.

\subsection{Model Architecture}
\label{subsec:architecture}

To bolster training on limited datasets, we utilised a pre-trained GeneformerV2~\cite{chen2024quantized} with $12$ self-attention blocks as the backbone of the encoder $G_{\text{E}}$. To avoid inimical training dynamics, we froze all but the final four self-attention modules, and appended a two-layer MLP to the \texttt{<cls>} token to project to the target dimensionality. A sigmoid activation function was used to ensure the projection reproduced the $[0, 1]$ range of the normalised spatial features. The default discriminator network $D$ was a $5$-layer MLP and $G_{\text{D}}$ was a $4$-layer MLP. As overfitting was observed on the significantly smaller brain dataset, we instead fully froze the backbone, and extended the projection head to three layers, while dropping a decoder layer. Throughout our experiments, we set $\lambda_{\text{ST}} = \lambda_{\text{SC}} = 1.0$ and $\lambda_{\text{GP}} = 10.0$.

\subsection{Baselines}
\label{subsec:baselines}

As a means of comparison, we trained and evaluated two other omics translation models, scACT \citep{xu2024scact} and scConfluence \citep{samaran2024scconfluence}, on the task of translating between \ac{scrnaseq} gene expression and spatially bulked expression. As both methodologies were not originally trained on this novel task, we made the following adaptations to enable a reasonable comparison, and trained each on the same data splits as SCXM.

\paragraph{scACT.} The \ac{atacseq} branch was replaced with one that used spatially bulked expression. CellTypist \citep{dominguez2022cross} annotations were used as the labels for the cell-type specific discriminator heads. \Ac{hvg} selection was not applied to the \ac{scrnaseq} input to maintain parity with SCXM.

\paragraph{scConfluence.} The inter-modality cost matrix required by the inverse optimal transport loss was computed via correlation distance on the log-normalised spatial panel $\mathcal{G}_{\text{ST}}$ in both the \ac{scrnaseq} and \ac{st} data. Binary cross-entropy was used as the reconstruction loss for the spatially bulked expression. \ac{hvg} selection was not applied.

%% file: sections/results.tex
\section{Results}
\label{sec:results}

\begin{figure}[]
  \centering
  \includegraphics[width=1.0\linewidth]{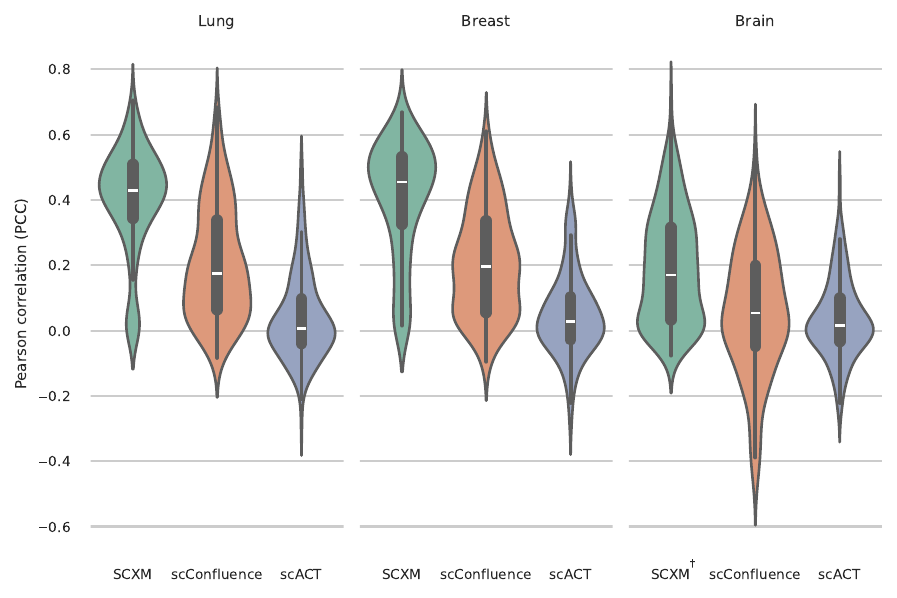}
  \caption{Gene-wise correlations on Xenium Prime validation data for cross-modal translation methods across three datasets. $\dagger$ On the smaller brain dataset, SCXM was trained with a frozen GeneformerV2 with MLP head.}
  \label{fig:crossmodal_results}
\end{figure}

\subsection{Cross-modal Translation}

SCXM outperformed prior methods for unpaired translation across all three curated datasets, as shown in Figure~\ref{fig:crossmodal_results}, with median \ac{pcc} of $0.4291$, $0.4551$, and $0.1700$ on the lung, breast, and brain datasets, respectively, compared with $0.1750$, $0.1964$, and $0.0573$ for scConfluence, and $0.0065$, $0.0284$, and $0.0151$ for scACT. We observed inferior performances on the significantly smaller brain dataset, although the best performing gene for SCXM scored a \ac{pcc} of $0.7085$, comparable with $0.7050$ and $0.6686$ in the lung and breast datasets. We postulate that the superior performance of SCXM for \ac{st} to \ac{scrnaseq} transfer may be attributed to its use of consistency penalties exploiting the count data relation between the modalities (Equations~\ref{eq:st_consistency} and~\ref{eq:sc_consistency}), as this amounts to semi-supervision, in contrast with the unsupervised cycle consistency of scACT and optimal transport plan of scConfluence.


\subsection{Niche Characterisation}

\begin{figure}[htbp]
  \centering
  \subfloat[Gene expression (binarised)\label{fig:five-a}]{%
    \includegraphics[height=0.43\linewidth]{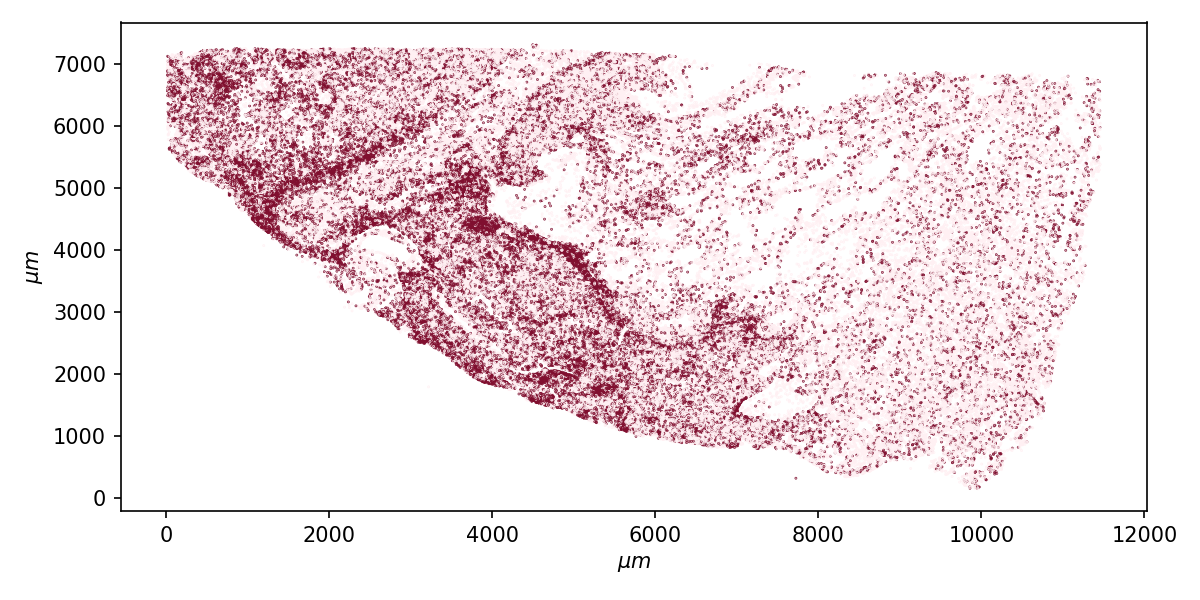}%
  }\\[1ex]
  \subfloat[Spatially bulked\label{fig:five-b}]{%
    \includegraphics[height=0.43\linewidth]{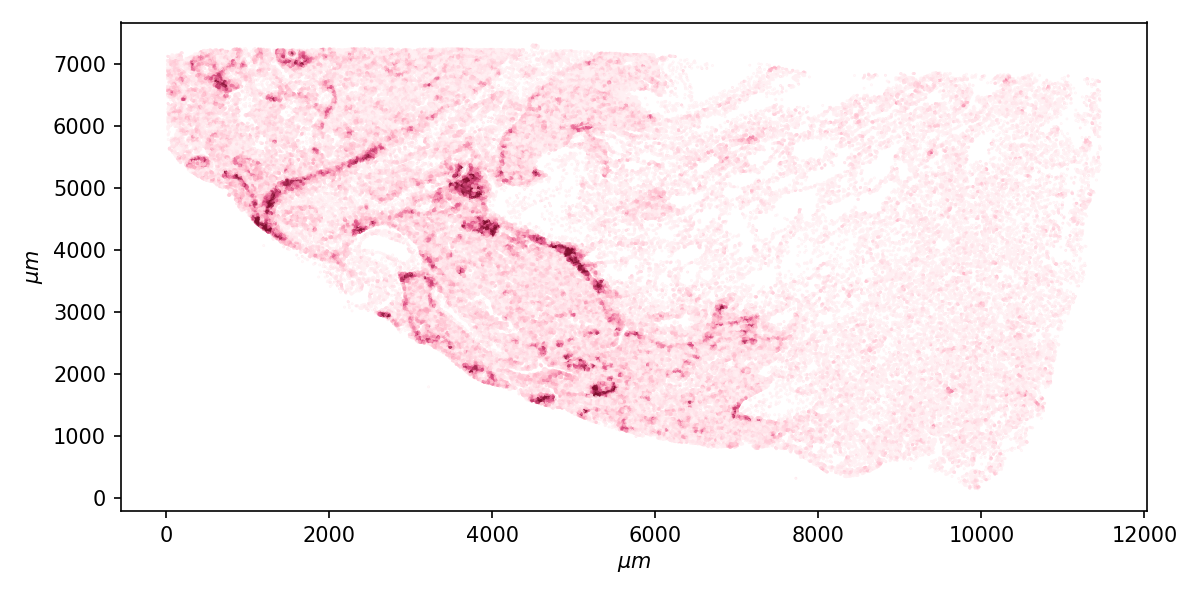}%
  }\\[1ex]
  \subfloat[Model prediction\label{fig:five-c}]{%
    \includegraphics[height=0.43\linewidth]{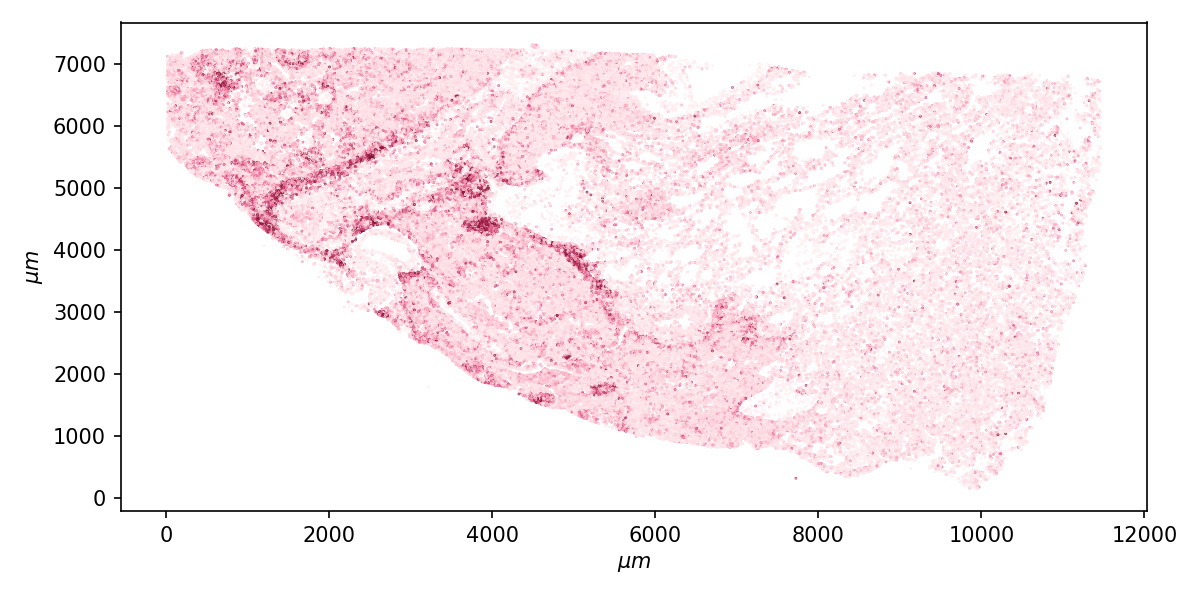}%
  }\\[1ex]
  \subfloat[Decoder output\label{fig:five-d}]{%
    \includegraphics[height=0.43\linewidth]{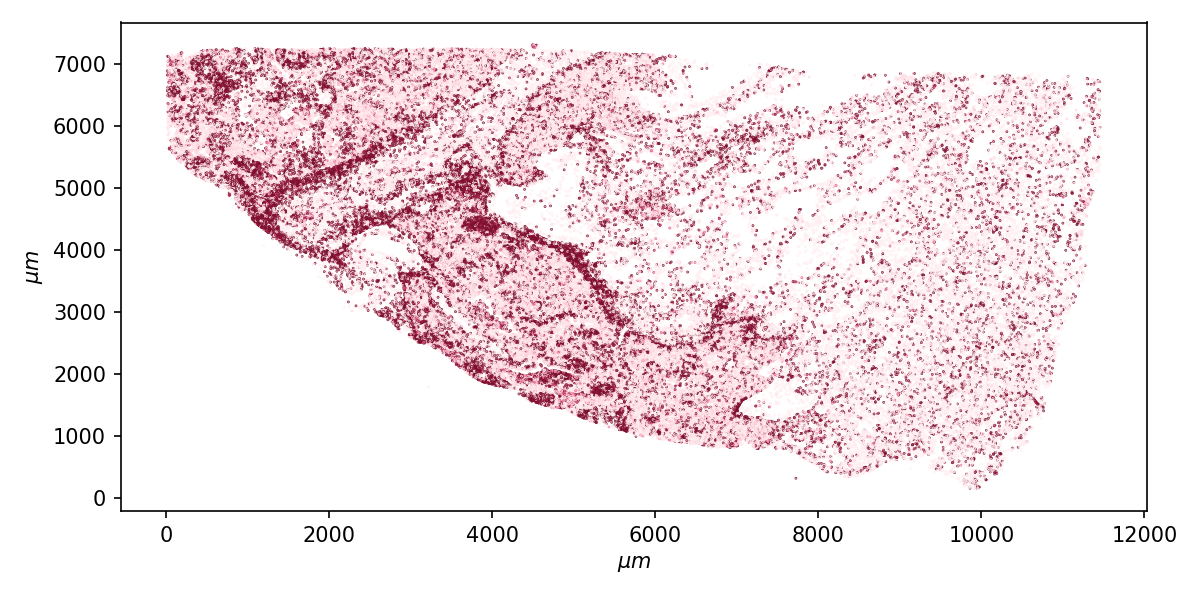}%
  }\\[1ex]
  \subfloat[Spatial subtyping\label{fig:five-e}]{%
    \includegraphics[height=0.43\linewidth]{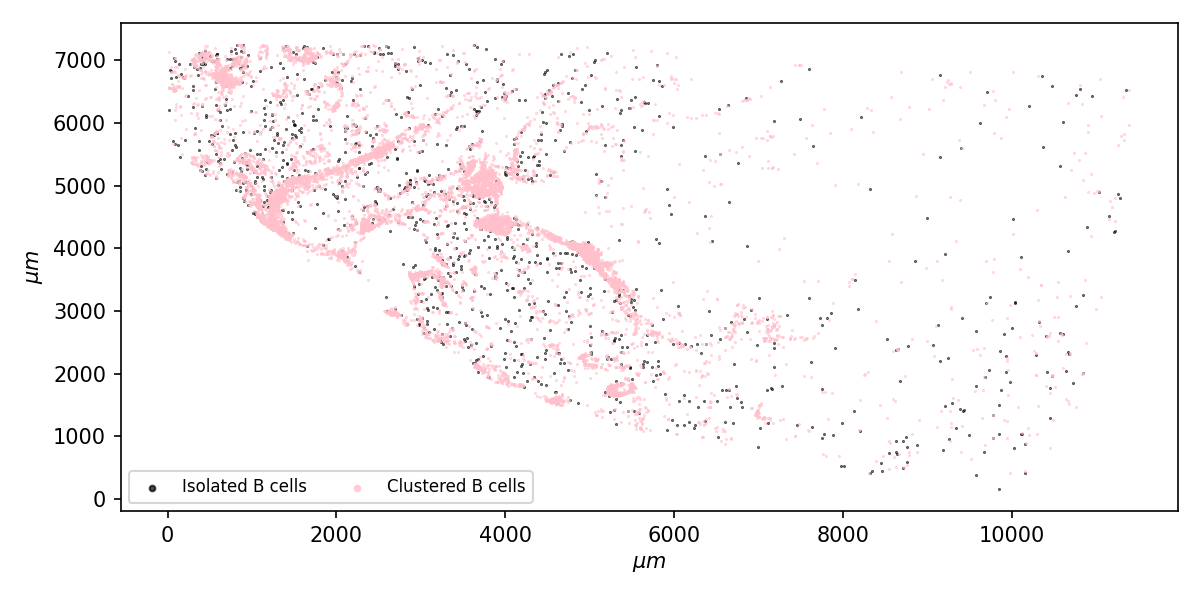}%
  }
  \caption{Whole-slide view of B cell spatial subtyping from \emph{CXCR4} gene expression input (a), target spatially bulked readout (b), model encoder prediction (c), model decoder output (d), and thresholded model outputs (e) as decision rule for distinguishing isolated B cells (black) and clustered B cells (pink).}
  \label{fig:five-panel-vertical}
\end{figure}

Predicting context as spatially bulked gene expression allows us to separate clustered expression from isolated expression, as shown in Figure~\ref{fig:foxj1} and Extended Figures~\ref{fig:marco}-\ref{fig:tnfrsf13c}. Such readouts may provide hints of the presence of niches in the tissue. We observe in Extended Figure~\ref{fig:cxcr4} an enrichment of B cells in regions predicted by SCXM to contain a high spatially bulked expression of \emph{CXCR4}. By thresholding the prediction at $0.5$, we can segment B cells across the lung tissue into isolated and clustered groups (Figure~\ref{fig:five-panel-vertical}). Applying differential analysis, we observe a >$4$ log-fold change for \emph{POU2AF1}, \emph{MS4A1}, and \emph{CD79A} ($p_{\text{adj}} < .001$, Benjamini-Hochberg). \emph{POU2AF1} in particular is implicated in the formation of biological niches such as germinal centres~\cite{tiedt2001ring}.


\subsection{Gene Set Enrichment Analysis}

In the lung dataset, we investigated whether SCXM predictions recapitulate certain aspects of spatial biology better than others. We selected a set of $25$ pathways relevant to lung tissue and subset to those genes which were measured within the spatial panel. A \ac{gsea} was performed over these pathways using \ac{pcc} as the ordinal metric, with a normalised enrichment score and Benjamini-Hochberg FDR correction used as in~\cite{Subramanian2005}. Results are shown in Table~\ref{tab:gsea}. The top two pathways, related to lung lineage and angiogenesis, were the most significant, highlighting the ability of SCXM to capture larger scale spatial biology. Six more borderline-significant pathways dealt largely with immune response, representing a more localised aspect of spatial biology at the epithelium-immune interface. Pathways related to cell-cell signalling and cell cycle were more difficult to recapitulate, perhaps due to fewer spatial constraints governing these processes. Thus, initial results indicate that the main strength of SCXM may intuitively lie in its ability to capture more prominent spatial context, and future work is planned to better understand its limits in this regard.

\input{tables/gsea}

%% file: tables/gsea.tex
\begin{table}[ht]
    \centering
    \caption{GSEA results for the lung dataset. NES: normalised enrichment score; FDR: false discovery rate (Benjamini-Hochberg).}
    \label{tab:gsea}
    \resizebox{\columnwidth}{!}{%
        \begin{tabular}{lrrrr}
            \toprule
            Pathway                    & Size/Overlap & NES       & $p$      & FDR   \\
            \midrule
            Lung lineage identity      & 50/15        & 1.835    & $<$0.001 & 0.003 \\
            Angiogenesis               & 36/18        & 1.894    & $<$0.001 & 0.003 \\
            T cell activation          & 100/26       & 1.328    & 0.010    & 0.063 \\
            Myeloid innate immunity    & 130/17       & 1.106    & 0.009    & 0.063 \\
            Interferon response        & 200/9        & 1.419    & 0.013    & 0.067 \\
            Immune checkpoint          & 80/12        & 1.373    & 0.025    & 0.099 \\
            Hedgehog signalling        & 47/7         & 1.431    & 0.028    & 0.099 \\
            Notch signalling           & 48/11        & 1.450    & 0.055    & 0.171 \\
            Integrin cell adhesion     & 199/13       & 1.276    & 0.074    & 0.206 \\
            B cell markers             & 71/13        & 1.173    & 0.092    & 0.229 \\
            Unfolded protein response  & 113/14       & 0.987    & 0.153    & 0.348 \\
            Chemokine signalling       & 190/10       & 1.136    & 0.185    & 0.386 \\
            PI3K/AKT/mTOR survival     & 354/15       & 0.976    & 0.316    & 0.607 \\
            EMT                        & 200/21       & 1.224    & 0.340    & 0.607 \\
            Hypoxia response           & 200/16       & 0.958    & 0.457    & 0.761 \\
            EGFR signalling            & 85/12        & 0.977    & 0.563    & 0.880 \\
            FGF signalling             & 55/11        & 1.050    & 0.656    & 0.965 \\
            Apoptosis                  & 136/14       & 1.010    & 0.906    & 1.000 \\
            NF-$\kappa$B signalling    & 100/13       & 0.977    & 0.986    & 1.000 \\
            Cell cycle / proliferation & 124/13       & 0.988    & 0.762    & 1.000 \\
            Hippo / YAP-TAZ            & 163/13       & 0.970    & 0.957    & 1.000 \\
            KEGG NSCLC pathway         & 66/17        & 1.070    & 0.912    & 1.000 \\
            Oxidative stress / NRF2    & 100/11       & 0.976    & 0.780    & 1.000 \\
            TGF-$\beta$ signalling     & 94/17        & $-$0.856 & 0.952    & 1.000 \\
            WNT signalling             & 158/19       & $-$0.573 & 1.000    & 1.000 \\
            \bottomrule
        \end{tabular}}
\end{table}

%% file: sections/conclusion.tex
\section{Conclusion}
\label{sec:conclusion}

With SCXM we have demonstrated preliminary success at cross-modal translation of \ac{scrnaseq} and \ac{st} data from completely unpaired data sources. Performance varied with dataset size, and future extensions could explore training on larger datasets. Moreover, assembling suitable dataset combinations spanning three modalities for the training and validation of SCXM proved to be challenging. Even when correspondence exists between the high-level properties of the studied cohorts, there is non-negligible domain shift across the modalities that makes model optimisation and validation difficult. One solution could be to move to a pan-tissue, pan-modal dataset, such as in~\cite{tejada2025nicheformer}, which could allow for model generalisation beyond study-specific biases. A challenge for SCXM would then be to integrate the disjoint spatial gene panels, which can vary significantly by technology and tissue type. We thus posit that future spatial niche modelling with SCXM could incorporate spatial cell type composition in addition to, or instead of, spatially bulked gene expression, facilitating a more universal representation of spatial biology.

%% file: sections/appendix.tex
\section{Extended Tables}

A summary of the datasets used for training and validating SCXM is provided in Table~\ref{tab:datasets}.

\input{tables/data}

\section{Extended Figures}
\label{app:details}

\begin{figure}[t]
  \centering
  \subfloat[Cell types\label{fig:marco-types}]{%
    \includegraphics[width=0.48\linewidth]{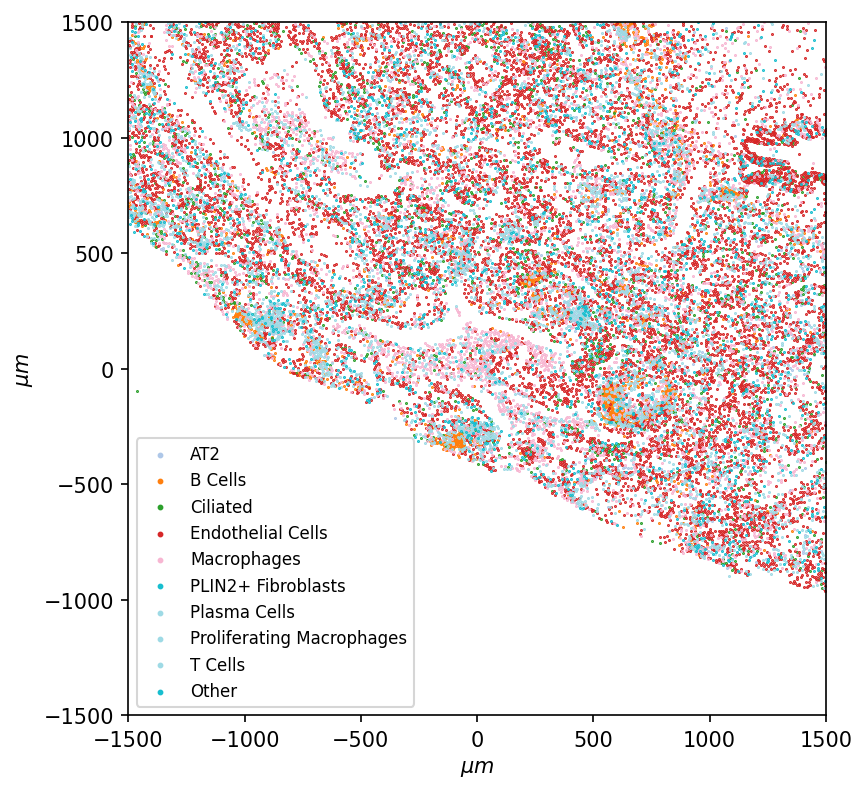}%
  }\hfill
  \subfloat[Gene expression (binarised)\label{fig:marco-exp}]{%
    \includegraphics[width=0.48\linewidth]{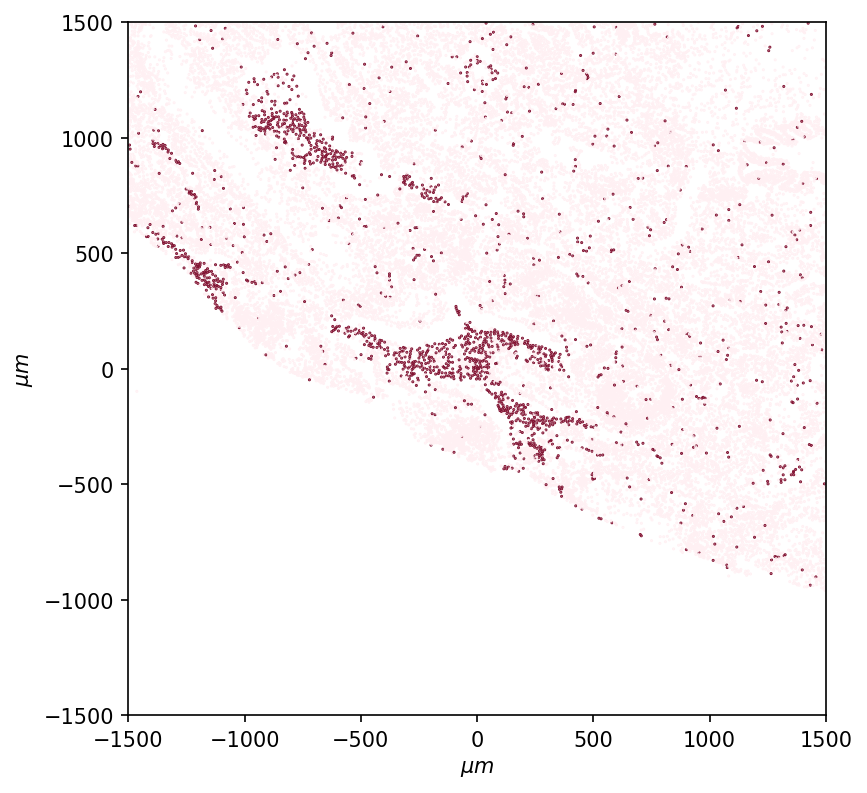}%
  }\\[1ex]
  \subfloat[Spatially bulked\label{fig:marco-conv}]{%
    \includegraphics[width=0.48\linewidth]{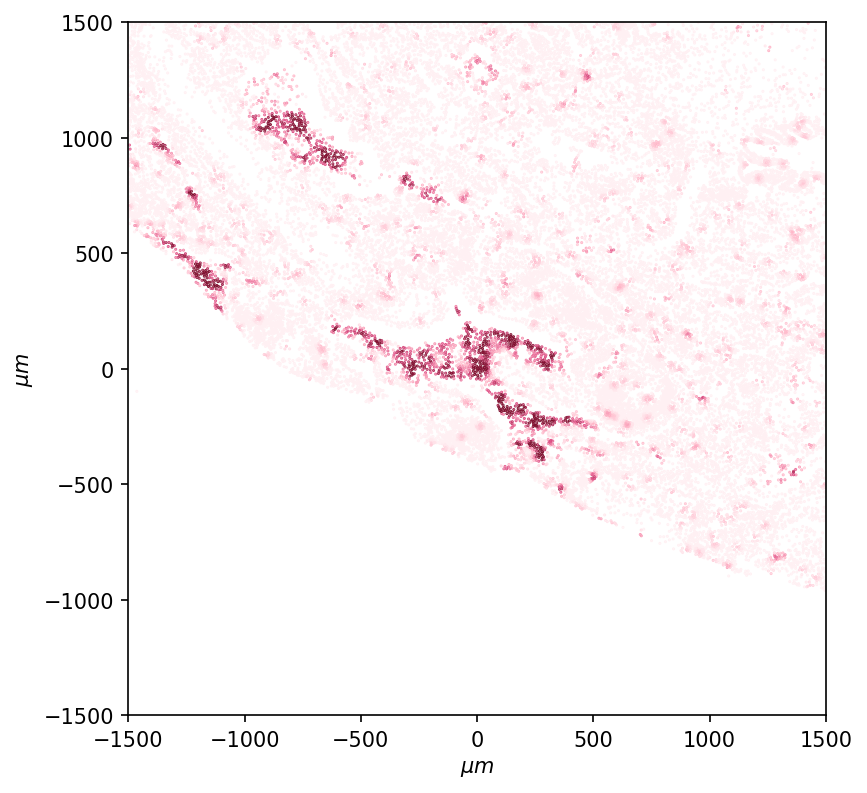}%
  }\hfill
  \subfloat[Model prediction\label{fig:marco-pred}]{%
    \includegraphics[width=0.48\linewidth]{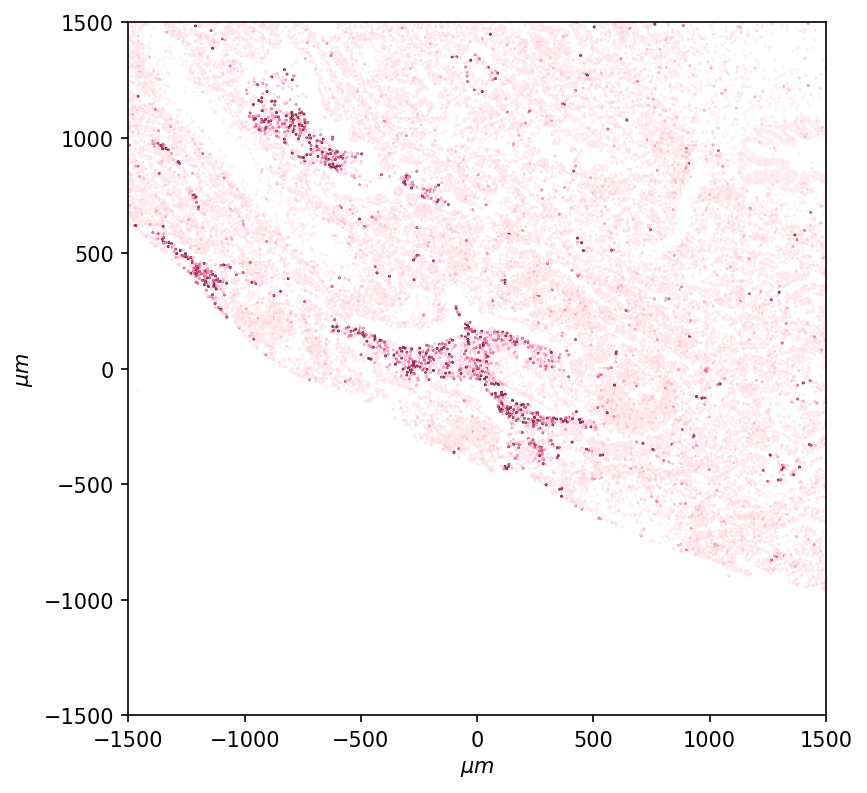}%
  }
  \caption{Sample region with cell types (a), gene expression inputs (b), targets (c) and model prediction (d) for \emph{MARCO} gene.}
  \label{fig:marco}
\end{figure}

\begin{figure}[t]
  \centering
  \subfloat[Cell types\label{fig:cxcr4-types}]{%
    \includegraphics[width=0.48\linewidth]{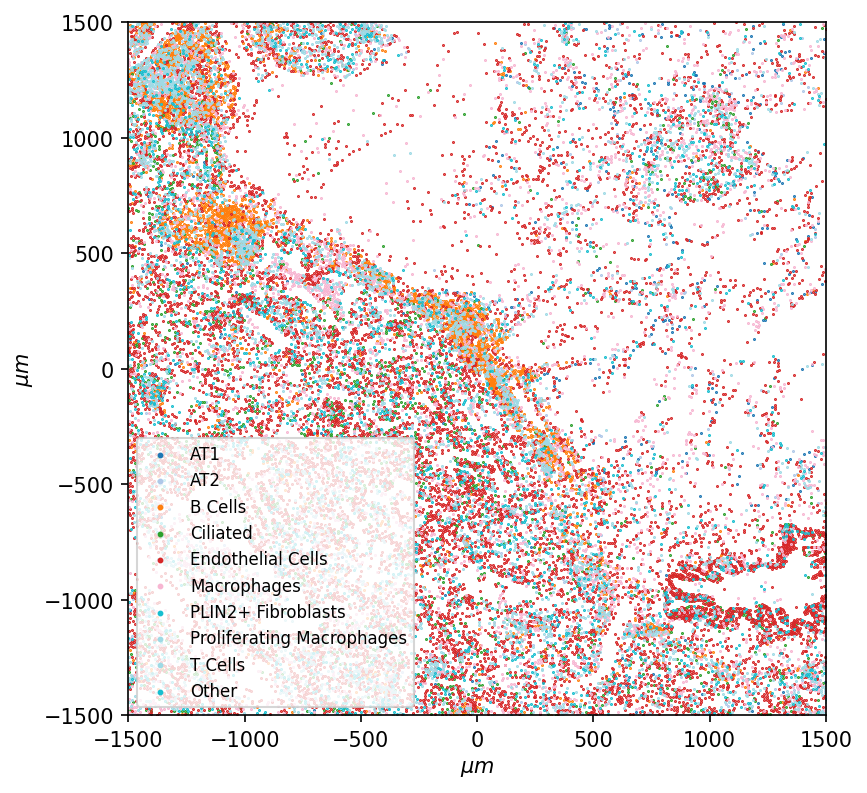}%
  }\hfill
  \subfloat[Gene expression (binarised)\label{fig:cxcr4-exp}]{%
    \includegraphics[width=0.48\linewidth]{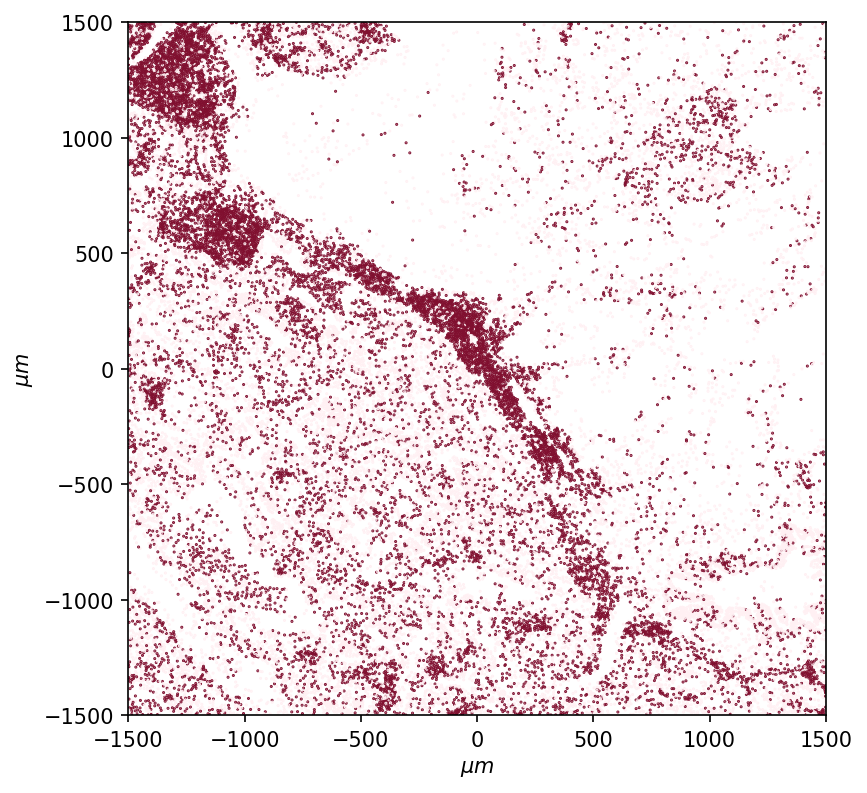}%
  }\\[1ex]
  \subfloat[Spatially bulked\label{fig:cxcr4-conv}]{%
    \includegraphics[width=0.48\linewidth]{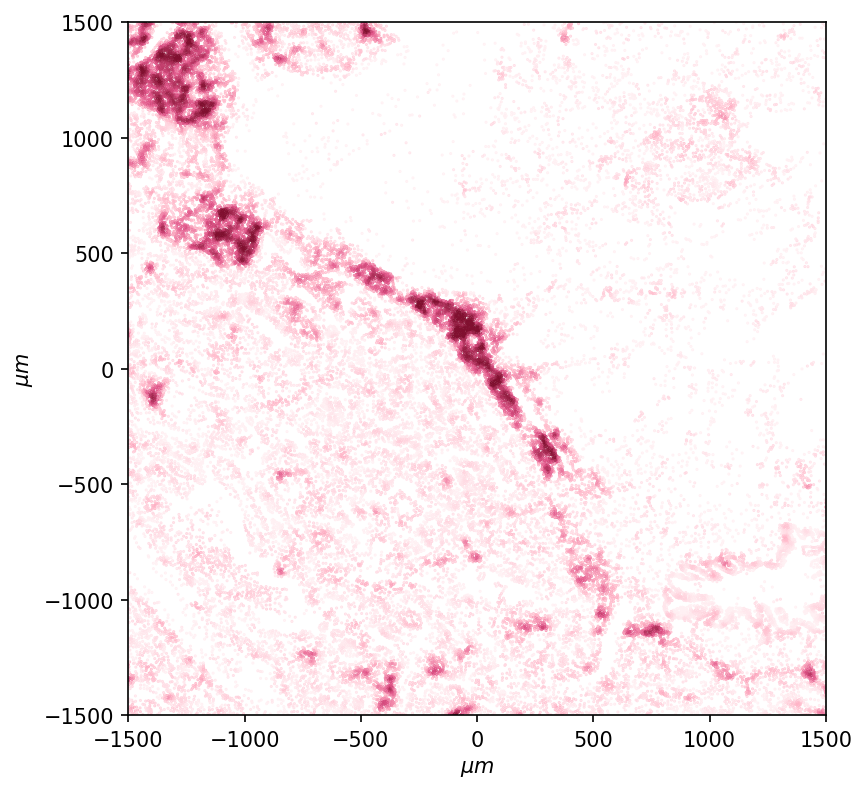}%
  }\hfill
  \subfloat[Model prediction\label{fig:cxcr4-pred}]{%
    \includegraphics[width=0.48\linewidth]{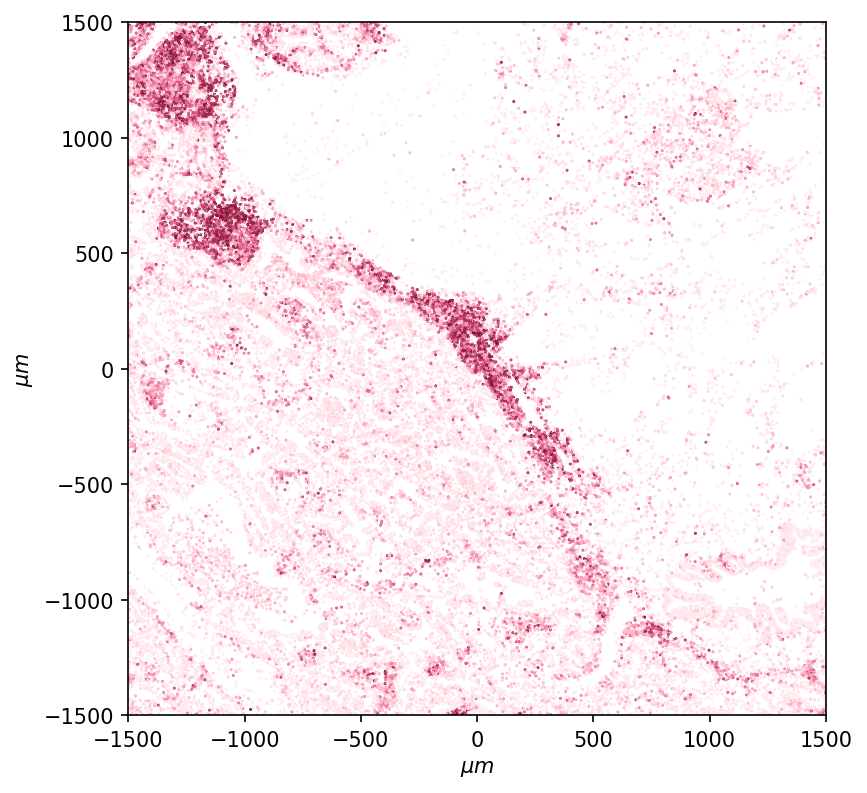}%
  }
  \caption{Sample region with cell types (a), gene expression inputs (b), targets (c) and model prediction (d) for \emph{CXCR4} gene.}
  \label{fig:cxcr4}
\end{figure}

\begin{figure}[t]
  \centering
  \subfloat[Cell types\label{fig:epcam-types}]{%
    \includegraphics[width=0.48\linewidth]{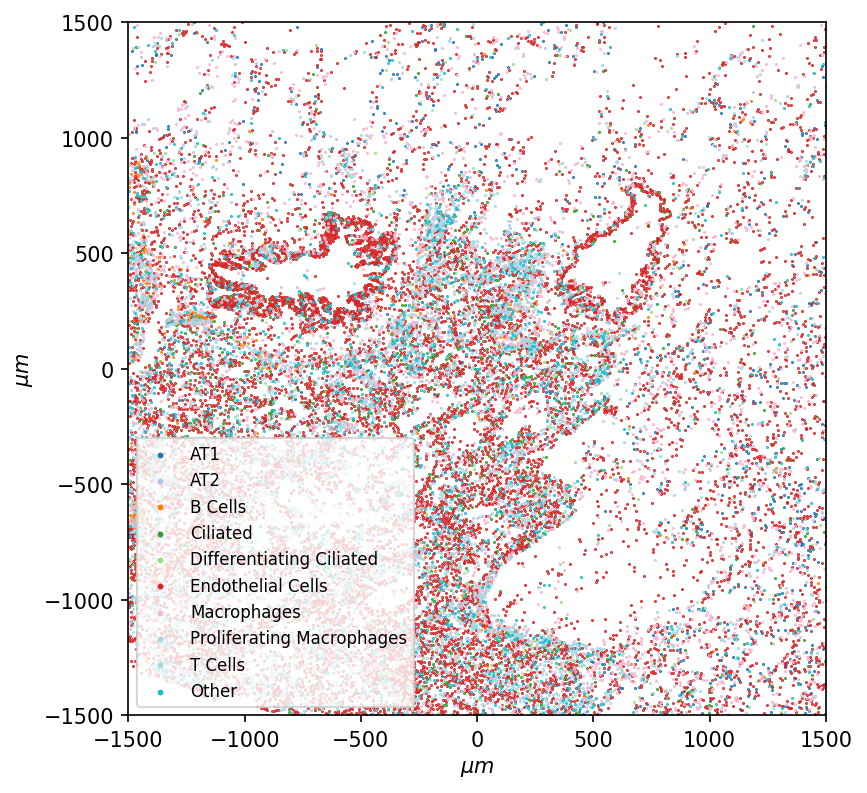}%
  }\hfill
  \subfloat[Gene expression (binarised)\label{fig:epcam-exp}]{%
    \includegraphics[width=0.48\linewidth]{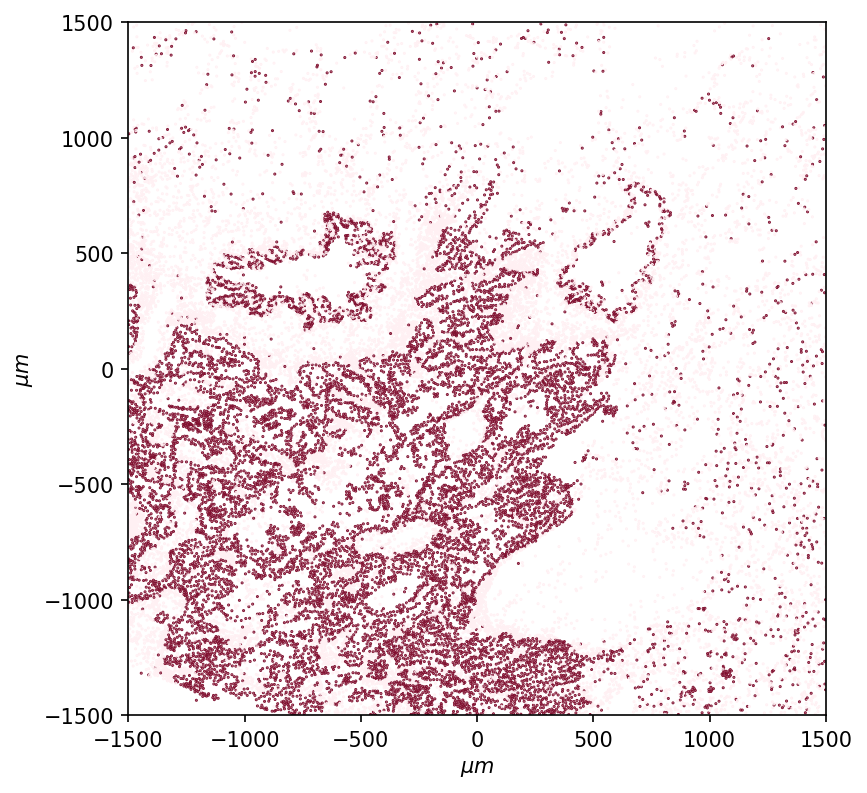}%
  }\\[1ex]
  \subfloat[Spatially bulked\label{fig:epcam-conv}]{%
    \includegraphics[width=0.48\linewidth]{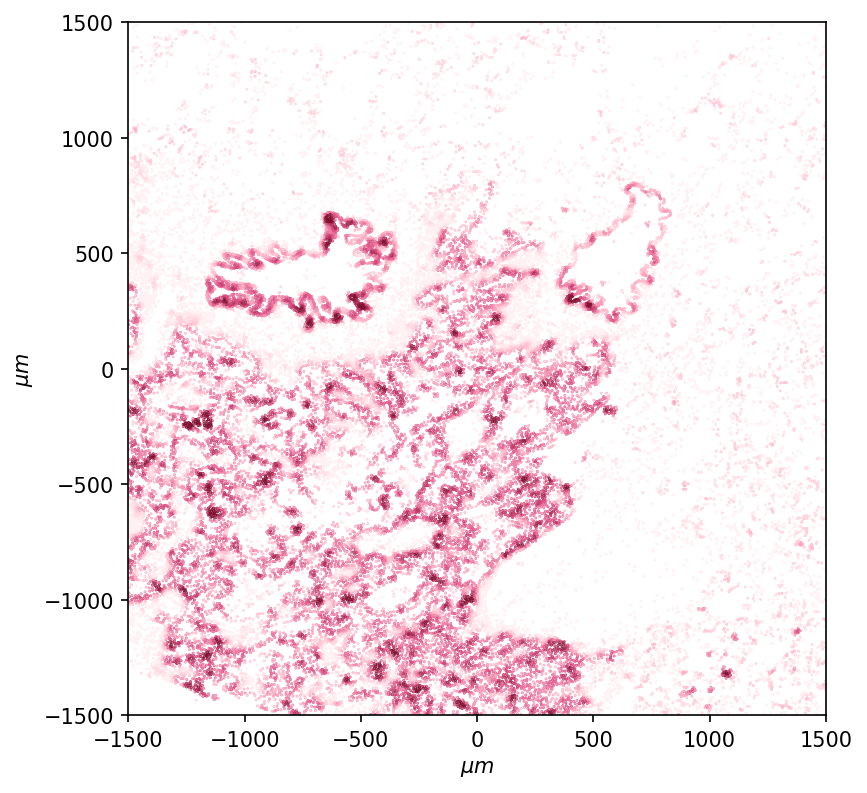}%
  }\hfill
  \subfloat[Model prediction\label{fig:epcam-pred}]{%
    \includegraphics[width=0.48\linewidth]{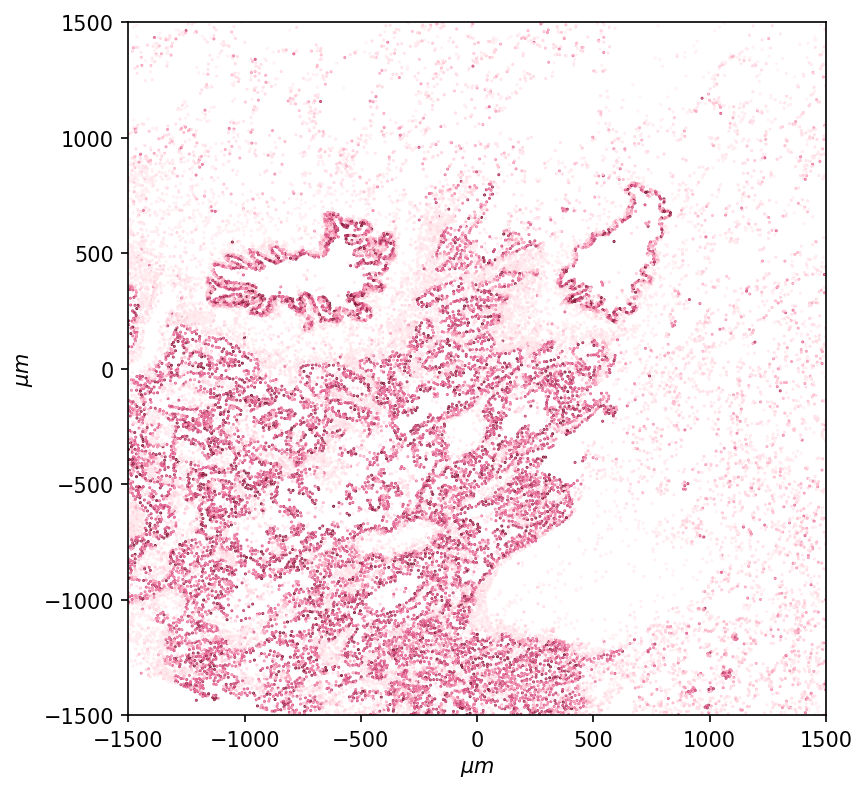}%
  }
  \caption{Sample region with cell types (a), gene expression inputs (b), targets (c) and model prediction (d) for \emph{EPCAM} gene.}
  \label{fig:epcam}
\end{figure}

\begin{figure}[t]
  \centering
  \subfloat[Cell types\label{fig:tnfrsf13c-types}]{%
    \includegraphics[width=0.48\linewidth]{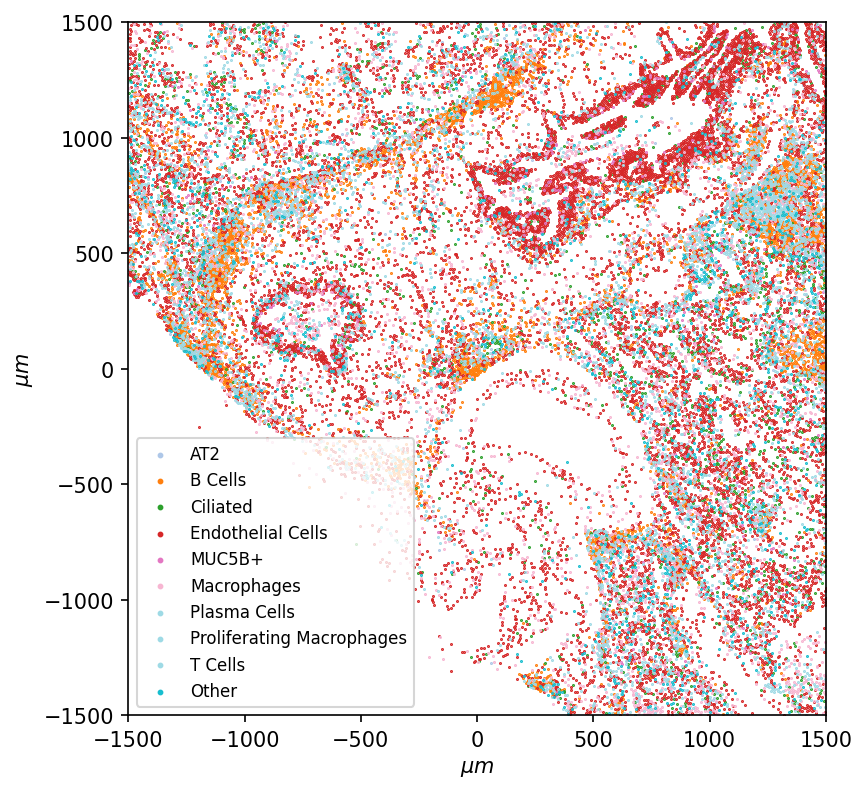}%
  }\hfill
  \subfloat[Gene expression (binarised)\label{fig:tnfrsf13c-exp}]{%
    \includegraphics[width=0.48\linewidth]{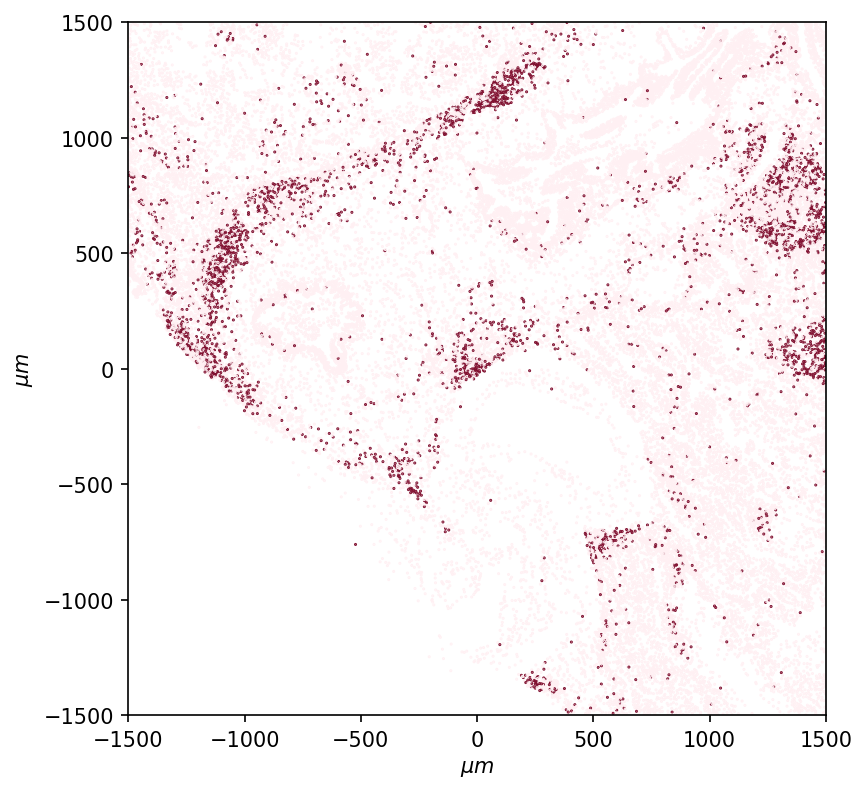}%
  }\\[1ex]
  \subfloat[Spatially bulked\label{fig:tnfrsf13c-conv}]{%
    \includegraphics[width=0.48\linewidth]{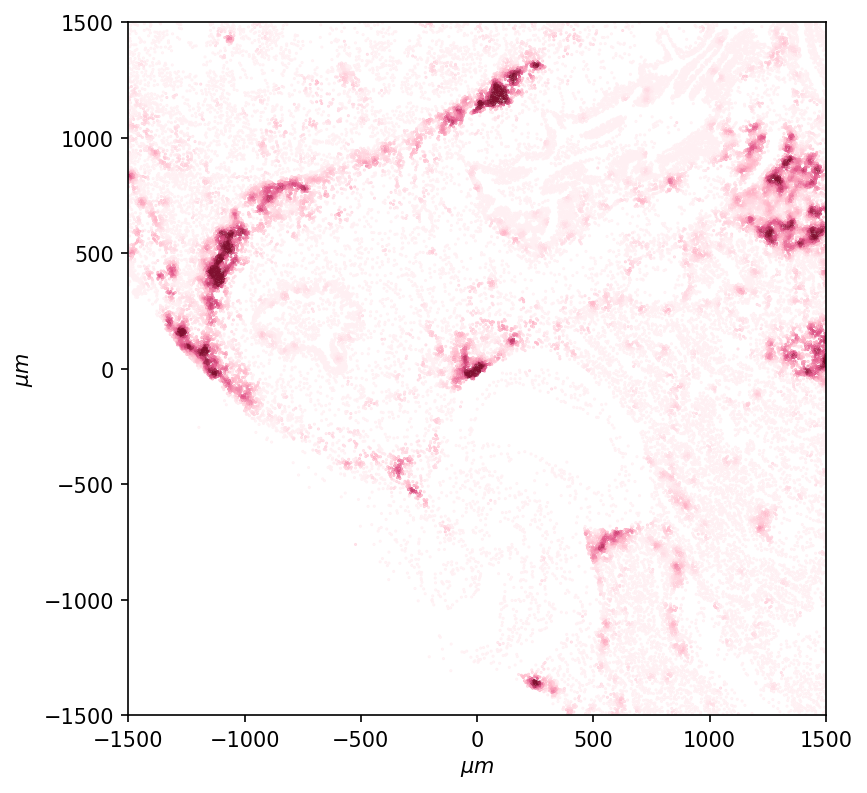}%
  }\hfill
  \subfloat[Model prediction\label{fig:tnfrsf13c-pred}]{%
    \includegraphics[width=0.48\linewidth]{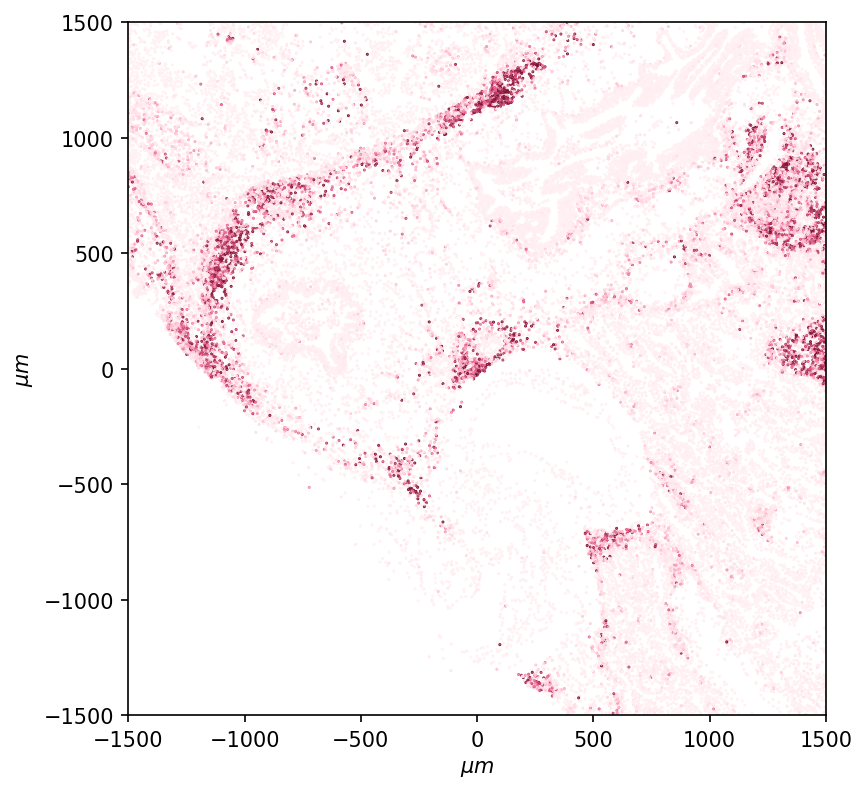}%
  }
  \caption{Sample region with cell types (a), gene expression inputs (b), targets (c) and model prediction (d) for \emph{TNFRSF13C} gene.}
  \label{fig:tnfrsf13c}
\end{figure}

Figures~\ref{fig:marco},~\ref{fig:cxcr4},~\ref{fig:epcam}, and~\ref{fig:tnfrsf13c} provide illustrative examples of SCXM inputs, targets, and predictions. Regions of dense prediction often correspond to clusters enriched in a given cell type, for example macrophages (Figure~\ref{fig:marco}) and B cells (Figure~\ref{fig:cxcr4}).

%% file: tables/data.tex
\begin{table*}[htbp]
    \centering
    \caption{Summary of datasets used for training and validating SCXM.}
    \label{tab:datasets}
    \small
    \renewcommand{\arraystretch}{1.3}
    \newcolumntype{L}[1]{>{\raggedright\arraybackslash}p{#1}}
    \begin{tabular}{@{}llL{4.5cm}L{5.5cm}@{}}
        \toprule
        \textbf{Dataset} & \textbf{Modality} & \textbf{Title} & \textbf{Retrieved from} \\
        \midrule
        \multirow{3}{*}{Lung (human)}   
            & scRNA-Seq    & An integrated cell atlas of the lung in health and disease~\cite{sikkema2023integrated}
            & \footnotesize\url{https://cellxgene.cziscience.com/collections/6f6d381a-7701-4781-935c-db10d30de293} \\[0.5ex]
            & Xenium v1    & Spatial transcriptomics identifies molecular niche dysregulation associated with distal lung remodeling in pulmonary fibrosis~\cite{vannan2025spatial}
            & \footnotesize\url{https://www.ncbi.nlm.nih.gov/geo/query/acc.cgi?acc=GSE276945} \\[0.5ex]
            & Xenium Prime & Post-Xenium Technical Note: Xenium v1 and Xenium Prime 5K for FFPE Human Lung Cancer~\cite{technical-note-prime-lung}
            & \footnotesize\url{https://www.10xgenomics.com/datasets/xenium-human-lung-cancer-post-xenium-technote} \\
        \midrule
        \multirow{3}{*}{Brain (mouse)}  
            & scRNA-Seq    & Mouse Brain Nuclei Isolated with Chromium Nuclei Isolation Kit~\cite{data-highlights-mouse-brain}
            & \footnotesize\url{https://www.10xgenomics.com/datasets/mouse-brain-nuclei-isolated-with-chromium-nuclei-isolation-kit-saltyez-protocol-and-10x-complex-tissue-dp-ct-sorted-and-ct-unsorted-1-standard} \\[0.5ex]
            & MERFISH      & Molecular and spatial signatures of mouse brain aging at single-cell resolution~\cite{allen2023molecular}
            & \footnotesize\url{https://cellxgene.cziscience.com/collections/31937775-0602-4e52-a799-b6acdd2bac2e} \\[0.5ex]
            & Xenium Prime & Fresh Frozen Mouse Brain Hemisphere with 5K Mouse Pan Tissue and Pathways Panel~\cite{technical-note-prime}
            & \footnotesize\url{https://www.10xgenomics.com/datasets/xenium-prime-fresh-frozen-mouse-brain} \\
        \midrule
        \multirow{3}{*}{Breast (human)} 
            & scRNA-Seq    & A highly resolved integrated single-cell atlas of human breast cancers~\cite{chen2026highly}
            & \footnotesize\url{https://cellxgene.cziscience.com/collections/9432ae97-4803-4b9f-8f64-2b41e42ad3cb} \\[0.5ex]
            & Xenium v1    & Biomarker Quantification in Breast Cancer using Xenium In Situ~\cite{janesick2025biomarker}
            & \footnotesize\url{https://www.10xgenomics.com/datasets/xenium-ffpe-human-breast-biomarkers} \\[0.5ex]
            & Xenium Prime & FFPE Human Breast Cancer with 5K Human Pan Tissue and Pathways Panel plus 100 Custom Genes~\cite{technical-note-prime}
            & \footnotesize\url{https://www.10xgenomics.com/datasets/xenium-prime-ffpe-human-breast-cancer} \\
        \bottomrule
    \end{tabular}
\end{table*}